\documentclass[
]{ceurart}

\usepackage{listings}
\usepackage{todonotes}
\usepackage{hyperref}

\begin{document}

\copyrightyear{2026}
\copyrightclause{Copyright for this paper by its authors.
                 Use permitted under Creative Commons License Attribution 4.0
                 International (CC BY 4.0).}

\conference{Joint Proceedings of REFSQ-2026 Workshops, Doctoral Symposium, Posters \& Tools Track, and Education and Training Track. Co-located with REFSQ 2026. Poznan, Poland, March 23-26, 2026}

\title{Experience Report on the Adaptable Integration of Requirements Engineering Courses into Curricula for Professionals}

\author[1,2]{Oleksandr Kosenkov}[
orcid=0000-0002-9971-1130,
email=oleksandr.kosenkov@bth.se,
url=https://regulatory-re.com/
]

\author[1,2]{Konstantin Blaschke}[
orcid=0009-0009-2095-9060,
email=blaschke@fortiss.org
]

\author[1]{Tony Gorschek}[
orcid=0000-0002-3646-235X,
email=tony.gorschek@bth.se
]

\author[1]{Michael Unterkalmsteiner}[
orcid=0000-0003-4118-0952,
email=michael.unterkalmsteiner@bth.se
]

\author[1]{Oleksandr Adamov}[
orcid=0000-0002-0120-5388,
email=oleksandr.adamov@bth.se
]

\author[1]{Davide Fucci}[
orcid=0000-0002-0679-4361,
email=davide.fucci@bth.se,
]

\address[1]{Blekinge Institute of Technology,
            Valhallavägen 10, 371 79 Karlskrona, Sweden}
\address[2]{fortiss GmbH,
            Guerickestr. 25, 80805, Munich, Germany}

\begin{abstract}
There is a growing demand for software engineering education (SEE) for professionals because of the increasing demand, active evolution of the technological landscape, and changes in the skills required by the practice. Integrating requirements engineering (RE) courses into SEE curricula for professionals systematically and effectively is challenging. In particular, curricula for professionals have different demands, are more dynamic, and modular in nature. In this study, we report on our experience in the development of three SEE curricula for professionals and the integration of RE courses into such curricula. We suggest basic principles for such integration and describe the systematic approach focused on course content mapping that we have developed. 
\end{abstract}


\maketitle
\section{Introduction}
Requirements engineering (RE) is often considered as one of the core subjects in engineering curricula~\cite{nwokeji2018panel} or relevant in related disciplines (e.g., business)~\cite{dekhtyar2020teaching}. In practice, incorporation of RE courses into software engineering (SE) curricula is done as an afterthought~\cite{hertz2018requirements, daun2023systematic} with prioritization of breadth over depth~\cite{moravanszky2025challenges}, and basic knowledge over industry-ready skills~\cite{daun2023systematic}. There is a persistent gap between what is taught in university courses and what is expected by industry ~\cite{hertz2018requirements, moravanszky2025challenges} despite the efforts to address it.

Existing models of formal higher education provide the required basic knowledge often without focusing on practical skills or new technological trends~\cite{burns2005graduate,dobslaw2023gap}. Herewith, practitioners resort to different forms of informal SE education ranging from Massive Open Online Courses to hackathons~\cite{nandi2016hackathons} or even meetups~\cite{ingram2020software}. In recent years, formal education and research institutions have been attempting to catch up with the informal education trends and develop their own courses and curricula for professionals.

However, the demand for curriculum for professionals (CfP) responding to dynamically changing learners and industrial needs is challenging the traditional approaches to curriculum development~\cite{vreuls2022responsive}. The importance of the contribution of instructors into the development of CfP is apparent, however, their engagement is constrained with the absence of suitable frameworks that among other would remove organizational barriers and clarify the roles in the process of curriculum development~\cite{vreuls2022responsive}.

We suggest that embedding RE courses into such curricula for professionals could be one way to address the industry-academia gap in RE education. From a pedagogical perspective, the peculiarities (e.g., curricula dynamics) of such newly emerging SE curricula for professionals are not well explored, and integration of RE courses into such curricula does not meet new challenges. In this study, we report our experience with the three cases of the development of SEE curricula for professionals and the integration of RE courses into such curricula. We focus on two research questions that we explored through our practice:

\begin{enumerate}
    \item[RQ1:] \textit{What characteristics of curricula for professionals (CfP) should be considered for integrating RE courses?}
    \item[RQ2:] \textit{How RE courses can be effectively integrated into curricula for professionals (CfP)?}
\end{enumerate}

Our experience shows that characteristics of CfP, such as content dynamics and autonomy of course instructors, should be taken into account while integrating the RE courses. Some of the peculiarities also provide opportunities for a better integration and improvement of RE teaching. To that end, we describe a content item-based approach for integrating RE courses that we used and developed across the three cases of CfP. Our study aims to amend the existing studies on teaching RE to professionals (e.g.,~\cite{hertz2018requirements, epifanio2023identifying}), and on RE courses in SEE curriculum (e.g.,~\cite{schlingensiepen2014competence,mohan2011teaching}) with (1) characteristics of CfP relevant for RE courses integration, and (2) a lightweight approach for integration of RE courses into CfP.

\section{Background}
\textit{Curriculum} is an essential feature of formal education and constitutes all the planned and guided learning experience of students~\cite{kelly2009curriculum}. It defines why, what, when, where, how and with whom students learn~\cite{braslavsky2003curriculum}. We define \textit{education curriculum for professionals (CfP)} as guided, consistent, coordinated learning experience of professionals across the courses towards the learning goals. We suggest that in education for professionals, curriculum should be developed independently of study duration or format (which can range from a course with a couple of modules to degrees) to provide synergy between courses and/or modules. Coordination of students' learning across courses can be achieved through their alignment~\cite{kopera20082, harden2001amee} at two levels. First, \textit{constructive alignment}~\cite{biggs1996enhancing} (or \textit{internal alignment}~\cite{shaltry2020new}) of learning outcomes, learning activities, and assessment on the level of a single course. Second, \textit{curriculum alignment} (or \textit{external alignment}~\cite{shaltry2020new}) of different courses at the level of the whole curriculum. In traditional education, curriculum alignment is conducted during planning and development. However, there is a demand to assure continuity of alignment~\cite{adeyemi2023course}, beyond the one-time alignment during the curriculum development~\cite{witham2021supporting}, look beyond the three traditional internal alignment factors (learning outcomes, activities, and assessment)~\cite{proitz2023consistency}, and make curriculum alignment more collaborative. Curriculum alignment can be instrumental to identify the contribution of courses to the achievement of curriculum goals~\cite{kopera20082, wijngaards2018improving}, analyze the curriculum as a whole~\cite{wsu2022quick}, establish the suitable sequence of courses~\cite{kopera20082}, and facilitate communication and collaboration between instructors~\cite{sumsion2004identifying}. In this paper, we consider RE \textit{course integration} as the process of alignment of RE course with the other courses in the CfP for assuring consistency of curriculum.
It is important to consider the integration of RE courses due to specific features and challenges of RE education, such as inherent uncertainties of RE~\cite{macaulay1995requirements}, or need for balancing theory instruction and process exposure~\cite{daun2023systematic}.

\section{Methodology}
We report the experience of the first author participating in all three projects described next (PROMIS, Software4KMU, and TASTE, see details in Section~\ref{sec:cases}) and other authors participating in one of the projects on education for professionals. First, in the PROMIS project, the first author of the study conducted interviews with the instructors of existing courses linked to the newly developed RE course. After developing the initial version of the alignment approach and developing a corresponding course, we conducted a focus group with PROMIS instructors and administrators to discuss the applicability and potential benefits of course alignment. Next, in the project Software4KMU, the first author of the study used the first version of the approach to consult the project manager on the development of the curriculum structure. Through observations and interviews with the project manager, the first author of the study collected the feedback about the benefits of the content items-based approach for the development of CfP and integration of RE courses. In the TASTE project, the first author of the study, with the support of the second author, used previous experience to suggest a strategy for collaboration between different remote research partners for the development of the curriculum. Also, our approach was used to collaboratively develop a learning path constituting a part of the TASTE curriculum. This work resulted in the most recent version of our approach and corresponding guidelines for its application, which are available in the open dataset.

\section{Cases descriptions}\label{sec:cases}
Next, we summarize the cases in which we applied our lightweight approach for RE course integration.

\textit{\href{https://promisedu.se/}{Professional Master in Information Security (PROMIS)}} is a curriculum for software engineering and IT professionals with at least two years of industry experience. It is implemented by the Computer Science and Software Engineering Departments of Blekinge Institute of Technology (BTH) and supported by 35 industry partners providing feedback. PROMIS comprises 15 online, instructor-led courses delivered via Canvas by academic and industry instructors with the support of two coordinators. The curriculum covers current cybersecurity topics and was developed around a high-level vision rather than a fixed structure, allowing instructors flexibility in course design. Completion of the PROMIS courses is awarded with ECTS credits towards a master's degree for professionals. The first author developed the RE course "Security, Privacy, and Compliance" focusing on the alignment of the three topics.

\textit{\href{https://se-toolbox.info/}{Software4KMU}} project aimed to equip software engineers in small and medium enterprises with the knowledge and tools for RE and quality assurance. The curriculum was developed by two departments of the research institute fortiss GmbH with feedback from two industry partners. An RE course was developed along with a quality assurance course and aligned within a single learning roadmap. The courses use online self-administered learning through a custom interactive learning website. Curriculum development was coordinated by a senior RE researcher involving 3 researchers and 2 student assistants. The materials were based on the general RE body of knowledge, the existing Artifact Model for Domain-independent Requirements Engineering, and materials developed for teaching the model in a university course. Curriculum development involved the adaptation of existing research results and publications. Curriculum does not provide credits and/or certificates upon completion.

\textit{\href{https://transformations-hub-taste.de/}{The Transformation Hub Automotive Software Engineering (TASTE)}} project supports integration of modern SE methods in the German automotive industry, and includes the development of educational paths for SE professionals. Five German research institutions collaborate to develop educational paths covering advanced SE topics delivered through self-administered online learning. No specific curriculum was predefined in the project, and each research institute developed learning paths (element of CfP) according to their specialization. To assure consistency alignment within and between different learning paths was required. The first and second authors of the study integrated RE, model-based systems engineering (MBSE), and quality assurance (QA) teaching materials to form a learning path for the professional profile of an automotive systems engineer. At the time of writing, the TASTE curriculum was drafted. The curriculum does not provide any credits and/or certificates upon completion.

All three CfPs were connected to university teaching and research of the instructors involved in their development. The courses specifically developed for PROMIS were planned for transfer to the undergraduate university curriculum. In the Software4KMU and TASTE project, instructors were reusing their research results and materials developed for university teaching. The cases reported in this paper cover both integration of newly developed RE course with existing courses (as in PROMIS project), and synchronous development of RE course along with other developed courses (as in Software4KMU, and TASTE projects).

\section{Results}
On the basis of our experience, we report the \textbf{characteristics of CfP that should be considered for integrating RE courses (RQ1)}, which are as follows:

\textit{Higher modularity and loose coupling of modules in CfP courses structure} make the structure of CfP and courses in it different from traditional university curriculum because of the demand to provide more selectivity in the teaching process, and higher dynamics of the curriculum. This requires restructuring of materials in cases of transformation between CfP and university courses or research results.

\textit{Orientation on practice and practical needs}, rather than educational standards or model curriculum, makes it hard to choose a reference point for the development of the curriculum. There is no body of knowledge about practice and skills required by professionals that can be used for CfP development. In this context, identifying the demands for RE learning is even more challenging because, in practice, RE is not well segregated and is usually combined with other SE processes.

\textit{Higher dynamism and heterogeneity of content} are required to cover a wide variety of topics and respond to the changing needs of professionals. Courses must also be continuously adapted to any changes in professional practice happening while the curriculum is taught (e.g., emergence of new RE method in practice). The increased need to react to changes results in faster obsolescence and replacement of existing content or faster addition of complementary content (e.g., practical cases) that should be coherently connected to existing content. This CfP characteristic enhances the autonomy of instructors and minimizes centralized curriculum alignment.

\textit{Minimal preliminary centralized alignment of CfP}. Unlike traditional university curricula, CfP does not involve detailed in-advance top-to-bottom planning and alignment of courses, but only a general high-level vision for the courses. This approach is usually intentionally taken to allow more autonomy for course instructors. As a result, alignment of courses to form a coherent curriculum needs to be conducted in a bottom-up way, starting with individual courses, rather than a top-to-bottom way.

\textit{Higher instructor autonomy and ownership of courses} are pivotal for flexibility of CfP without bureaucratic constraints. In our experience, integrating the RE course was enabled by collaboration with other instructors and their recognition of how it supports their courses and CfP through prerequisite knowledge, reinforcement of material.

\textit{No need for complete curriculum alignment}. We suggest that integration of all courses into a single CfP is not required. It is more beneficial to integrate courses into a couple of interconnected CfP components (learning paths), each focusing on a particular practical skill set or role, which will form the CfP. Such learning paths, developed bottom-up, are more viable and important to guide heterogeneous students.

\textit{Assessment plays a secondary role}.
Content turned out to be pivotal for integrating RE courses in CfP. In our experience, learning activities and assessment received only marginal attention both in course alignment and in the CfP overall. This is related to the different purpose of CfP to provide practically applicable knowledge, rather than a formal qualification.
\\

Next, on the basis of our experience, we suggest the following \textbf{recommendations for effectively integrating RE courses into CfP (RQ2)}.

\textit{Use shared artifacts that involved instructors can jointly use} to specify information about their courses (e.g., course module descriptions, specifications of learning objectives). This enables shared understanding among instructors and opportunities to work asynchronously. Making the artifacts useful and usable can enable updateability of alignment and eventually continuous alignment between RE and other courses.

\textit{Make integration instructor-driven, rather than administration-driven} to identify the degree and topics for alignment. This enables meaningful connections between RE and other courses, and better contextualization within CfP.

\textit{Use course content as a pivot for RE course integration and alignment} because it is the most tangible and convenient for instructors to focus on. Other alignment aspects (learning outcomes, activities, and assessment) can be effectively considered on the basis of content.

\textit{Contextualize RE courses within CfP}. The RE course should be adjusted to better fit at least the courses it is most related to. This helps to make the RE course more grounded and concrete for students and builds direct connections to more practical hands-on skills. For example, in the PROMIS project, threat modeling was added to the RE course after reviewing the related courses. This allowed us to strengthen the teaching of threat modeling across different courses.

\textit{Assure flexibility and adaptivity of RE course} content and form during the integration and in the future. Demonstrating adaptivity is required when collaborating with instructors and industry partners.

\textit{Transcend the traditional course boundaries to achieve better practice orientation}. In TASTE, we merged RE, MBSE, and QA courses to form a single role-specific learning path of automotive software engineer, along with other learning paths forming CfP. Such a learning path is more understandable and convenient to practitioners than separate courses. Such an approach, however, requires ensuring the manageability of the merged courses on the instructors' side.

\textit{Reuse the learning materials from other courses in the RE course}. Reusing content across the courses should allow consideration of real-world cases from multiple perspectives, including the RE perspective.

\textbf{Suggested integration approach (RQ2)}.
Next, we describe the most recent version of our approach to the integration of RE courses into CfP, which was applied in the TASTE project and previous versions of which were used in PROMIS and Software4KMU projects. The steps of the approach are as follows.

\textit{Identifying courses related to RE}.
As the first step, existing courses or a vision for courses in a CfP must be analyzed to identify the most related courses with which RE course should be integrated. This is done by analyzing learning outcomes, learning activities, assessment methods, and course content.

\textit{Mapping of courses content}. Next, the content of the RE course and existing or newly developed courses with which it will be integrated is divided into \textit{content items (CI)}---units of related content learning of which is estimated to be not longer than 10-15 minutes. The identification of CIs can be driven by the availability of the content for a new course, vision of CfP, or other considerations. Each content item should be specified in a shared document used to establish a common understanding during integration. The specification of each CI should contain a title reflecting the main content and a description containing the main topics or concepts covered in the item. In the TASTE project, we used the draw.io tool and UML class notation, allowing us to specify the CI name as a class instance name and the taught topics and concepts as class properties. Next, two strategies can be applied. In the first strategy, more suitable for the integration of RE courses with existing courses, CIs for the RE and other courses are put in sequence, in which they should be learned for each course independently. After sequencing is completed for each course, the sequenced CIs are brought together for collaborative alignment and integration (see the next paragraph). In the second strategy, applicable when new RE and other new courses are developed concurrently, the content of each course is specified in CIs; they are brought for collaborative consideration by the instructors of the RE and related courses (sequencing of CIs for each course independently can be skipped). Mapping can be conducted by each instructor for their corresponding courses (preferably) or by the RE course instructor for all the mapped courses.

\textit{Collaborative CIs' organization}. As soon as all the CIs are specified, the instructor of the RE course starts to collaborate with instructors of the related courses to identify intersections in titles or topics and concepts of CIs. If any intersection is identified, the suitable order for the CIs is identified, and they are put in a single sequence. If there is no vision towards course integration within CfP, sequencing is used as a tool for exploration and bottom-up derivation of potential learning paths, CfP, and identification of learning outcomes (see the next paragraphs). If there is a vision towards the integration of the courses and CfP overall, sequencing of the courses can be guided by such a vision and learning outcomes defined in a top-down way for a learning path (e.g., gaining knowledge required for a specific role) or CfP. The process of sequencing the CIs is the main part of the alignment between the courses. Herewith, changes to the CIs should be made in order to strengthen the alignment between the courses. In the TASTE project, collaborative work on the content items both helped to tailor the RE content items and also inspired the addition of the new ones.

\textit{Learning path(s) and modules specification}. As soon as CIs of all the courses are arranged in a single sequence, CIs can be organized into \textit{learning modules} (groups of closely related CIs on the same topic) and \textit{learning paths} (a sequence of content to be learned to progressively build knowledge progressively towards a professional goal/learning outcome within CfP). Learning modules are optional, but they can be useful to manage large and complex courses. They can be defined as obligatory learning modules (which form a part of the main flow of CIs) or optional learning modules taken by students on a voluntary basis, forming a parallel sequence to the main CIs sequence but usually connected to at least one CI in the main flow. Derivation of the learning path is the goal of the integration and is mandatory. It is achieved by arranging CIs in a sequence, providing a consistent learning experience. Derivation of learning modules and paths is iterative and can involve multiple interconnected changes in CIs across the courses (four versions of the learning path were developed in TASTE). Derivation and specification of the learning path can be performed both jointly (synchronously) or by each instructor individually (asynchronously). In the TASTE project, we have used both approaches, and especially enjoyed the convenience of effectively working together on the courses' alignment asynchronously.

\textit{Learning materials specification}. As soon as the learning path is stable, instructors can add \textit{material items} which describe concrete materials that will be used to teach the content described by CIs. Material items can enable alignment and reuse of the materials used across CfP (which was identified as important in the PROMIS project). For now, the description of material items is limited to their name, type of material (e.g., book, paper), and reference to the material location.

\textit{Learning outcomes specification or verification}. Finally, learning outcomes are specified for the created learning path (if required, for learning modules) on the basis of the specified and sequenced CIs. In case the aligned learning outcomes were already envisioned or specified (for each course, learning path, or CfP), their attainability should be verified towards the specified CIs. In the TASTE project, only joint aligned learning outcomes were specified at the level of the learning path. In PROMIS, aligned learning outcomes are planned to be specified for each learning path (no learning paths were developed at the time of writing), along with learning outcomes formulated for each independent course.

Depending on the depth of the alignment, the resulting aligned courses can be presented as one single integrated learning path (in TASTE) or as a compound learning path which can be presented as completely standalone (in PROMIS), but aiming to achieve outcomes of the same learning path. At the time of writing, learning paths for PROMIS curricula were planned for development. However, students of the RE course were presented with a mapping of relationships between courses on the basis of the alignment activities. Such mapping was appreciated by the students as useful to navigate different courses and the CfP overall. In the TASTE project, we have specified a learning path with 35 CIs combining RE course content (4 CIs in obligatory and 7 CIs in optional modules), MBSE course content (7 CIs in obligatory and 6 CIs in optional modules), and QA course content (8 CIs in obligatory and 3 CIs in optional modules). The learning path included 3 obligatory modules (2 modules composed of CIs from all three courses, and 1 module composed of CIs from a single course) and 5 optional modules, each having CIs belonging to one course only. The guidelines used for the development of the learning path and the learning path integrating RE, MBSE, and advanced SE courses developed in the TASTE project can be found in the open data set in the \href{https://doi.org/10.5281/zenodo.18305361}{Zenodo repository (DOI: 10.5281/zenodo.18305362)}.

\section{Conclusions}
Our suggested approach in all three cases enabled the integration of RE courses into CfP. In the PROMIS project, the RE course developed following the first version of our approach has been taught for three years and received positive feedback from the students. In the Software4KMU and TASTE project, students' feedback has not collected so far. However, the Software4KMU project manager positively assessed the suggested content item-based approach for CfP development, such as a modular structure enabling assignment and control of content development by different contributors. In the future, we plan to explore the viability of our approach from the perspective of both students, course instructors, CfP administrators, and industrial partners in a more structured way.

Our experience suggests that universities and research institutes need to approach the development of curriculum differently when targeting professionals. Integration of RE courses into the curriculum for professionals requires their adaptation and contextualization through content adjustments and the development of connections to other courses. Such contextualization allows for the assurance of practical orientation of the RE course (characteristic of CfP overall), supports the high autonomy of instructors, and fits the high dynamics and modularity of CfP and other courses.

We suggest that our approach has the potential to address such RE education-specific challenges as balancing theory and practically applicable skills, teaching narrow RE topics like security or safety RE~\cite{daun2023systematic}, or facilitating the domain adaptation of RE education~\cite{wohlgemuth2024domain} through better contextualization of RE courses within the curriculum. Applicability of this approach to other courses not related to RE remains in question, as they may not have a demand for in-depth alignment and contextualization.

Our experience, based on projects in two different countries (Sweden and Germany) and executed in two different types of institutions (a university and a research institute), suggests that our results can be transferred to other institutions developing CfP. However, we acknowledge that the approach requires further refinement for its successful transfer. Specifically, clear purposes and benefits for instructors to be engaged in the course alignment should be developed. Also, the interaction between instructors and administrative coordinators should be further clarified. Transferring our suggested approach can demand strong internal commitment and promotion to engage other instructors in the necessary activities. We plan to transfer our approach to institutions through project-based collaboration or long-term institutional partnership to support its adaptation and further development. We encourage colleagues interested in the approach or in our experience with integrating requirements engineering courses into the curriculum for professionals to contact the corresponding author with any questions or suggestions.

In this paper, we report the core aspects of our experience in integrating RE courses into CfP. Accordingly, we omit some aspects of CfP development that did not have a decisive impact on the integration of the RE courses, such as the involvement of external stakeholders or the interaction between course instructors and administrative course coordinators. However, we encourage other researchers to explore the impact of these aspects on RE course integration in their respective contexts. This study also does not analyze professional certifications and training programs (e.g., provided by the International Requirements Engineering Board) as a form of CfP, but rather leaves this for future work. We also plan to further examine and position our approach to course integration within the literature on didactics in future work.

We hope that our reported experience and approach can contribute to further developments in teaching RE to professionals and invite colleagues to further discussion on this matter.
\\

\section*{Contributions}
O. Kosenkov: Conceptualization; Methodology, Investigation; Data collection (lead); Original draft \& editing. K. Blaschke: Data collection; Draft reviewing \& editing. T. Gorschek: Data collection; Draft reviewing \& editing. M. Unterkalmsteiner: Supervision. O. Adamov: Data collection; Draft reviewing \& editing. D. Fucci: Data collection; Draft reviewing \& editing; Supervision.
\\

\section*{Acknowledgements}
The authors express their gratitude to Yevhen Ivanenko, Anton Luckhardt, Anna Eriksson, Monique Johansson, Oleksii Baranovskyi, Severin Kacianka, PROMIS and TASTE project teams for their support.
\\

\section*{Declaration on Generative AI}
During the preparation of this work, the author(s) used Grammarly, ChatGPT in order to: Grammar and spelling check, Paraphrase and reword. After using this tool/service, the author(s) reviewed and edited the content as needed and take(s) full responsibility for the publication’s content.

\bibliography{bibliography}
\end{document}